# Crossed-beam pump-probe microscopy

JUN JIANG,[1] WARREN S. WARREN,[2] AND MARTIN C. FISCHER[3,*]

[1]*Department of Biomedical Engineering, Duke University, Durham, North Carolina, 27708, USA*
[2]*Departments of Chemistry, Physics, Biomedical Engineering, and Radiology, Duke University, Durham, North Carolina, 27708, USA*
[3]*Departments of Chemistry and Physics, Duke University, Durham, North Carolina, 27708, USA*
*\*martin.fischer@duke.edu*

**Abstract:** We present a new imaging method for pump-probe microscopy that explores non-collinear excitation. This method (crossed-beam pump-probe microscopy, or CBPM) can significantly improve the axial resolution when imaging through low-NA lenses, providing an alternative way for depth resolved, large field-of-view imaging. We performed a proof-of-concept demonstration, characterized CBPM's resolution using different imaging lenses, and measured an enhanced axial resolution for certain types of low-NA lenses.

## 1. Introduction

Pump-probe microscopy is a specific implementation of nonlinear microscopy that can measure transient absorption processes like two-photon absorption, excited state absorption, stimulated emission, ground state depletion, and stimulated Raman scattering (Fig. 1(a)) for structural and functional imaging. Transient absorption can provide molecular specificity without the need for exogenous dyes and labels, and measure the electronic and vibrational dynamics of materials at high spatial and temporal resolution [1, 2]. Because of the good sensitivity and specificity, pump-probe microscopy has found applications in the fields of material science, biomedicine, and art conservation [3, 4].

To achieve a high sensitivity, pump-probe microscopy employs an amplitude modulation transfer scheme [1, 5], in which a modulated pump pulse train interacts with an unmodulated probe pulse train in the sample. Nonlinear interactions transfer some of the pump modulation onto the probe, where it can be detected by collecting reflected or transmitted probe light and demodulating the signal in a lock-in amplifier. In the most common pump-probe microscopy implementation, the pump and probe beams are overlapped collinearly and focused onto a common spot in the sample, which is then scanned in the region of interest (Fig. 1(b), [6]). With this scheme, the spatial resolution is determined by the effective numerical aperture (NA) of the objective lens, and can be estimated using the equations derived for two-photon excited fluorescence microscopy [7]:

$$r_{FWHM} \approx \frac{0.38\lambda}{NA}, \quad z_{FWHM} \approx \frac{0.63\lambda}{n - \sqrt{n^2 - NA^2}} \qquad (1)$$

Here, the lateral ($r_{FWHM}$) and axial ($z_{FWHM}$) resolution are defined as the full-width at half-maximum (FWHM) of the point-spread-function (PSF); $\lambda$ is the wavelength of the excitation light, $n$ is the refractive index of the immersion medium, and we assumed similar optical modes for pump and probe. An example of the resolution as a function of NA is plotted in Fig. 1(c), using $\lambda = 800\ nm$ and $n = 1$.

Although high-NA objective lenses can offer sub-micron resolution, they are restricted in applications that require a long working distance (WD), large field-of-view (FOV), or compact setup (e.g. electrophysiology experiments, miniaturized devices). This is because objective lens generally need to trade off large NA for a decreased WD or a larger size in order to retain good aberration correction. Recently, three strategies have been demonstrated in two-photon imaging to expand the FOV or WD of high-NA lenses: 1) place the scanning mirrors in the post-objective space [8]; 2) accept the large size objective lens [9-11]; or 3) utilize multiple objective lenses [12]. While these methods have demonstrated large imaging areas (FOV >3 × 3 $mm^2$) with decent resolution, they all have limitations. For example, post-objective scanning sacrifices the WD of the objective lens, while using multiple objectives is not scalable. Very large objectives add dispersion and do not lend themselves to portable microscopes.

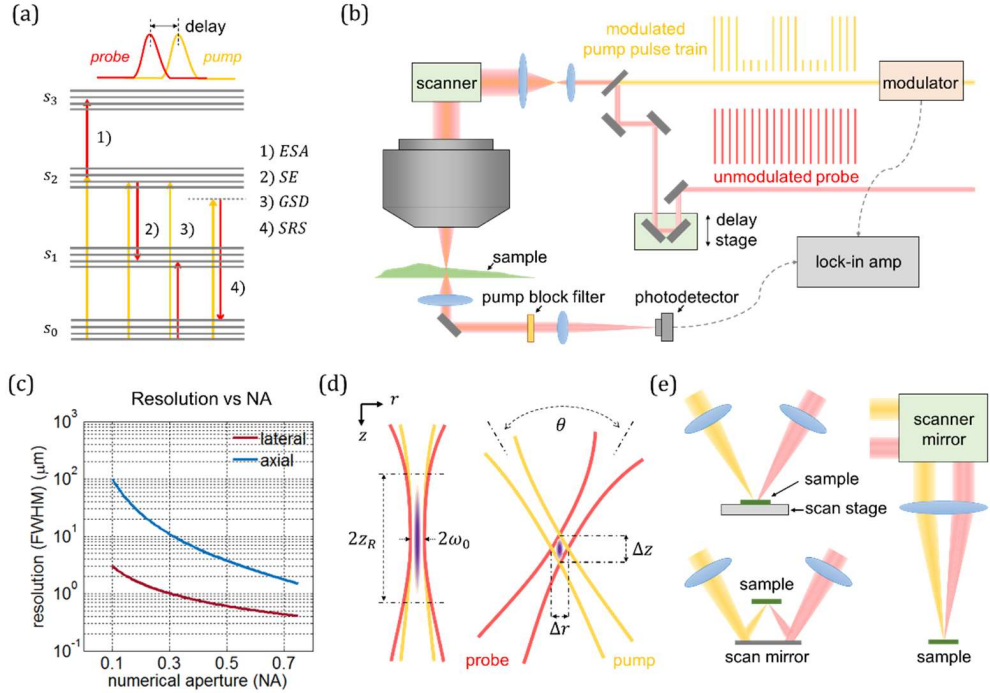

Fig. 1. (a) Some accessible transient absorption processes in pump-probe microscopy. ESA: excited state absorption, SE: stimulated emission, GSD: ground state depletion, SRS: stimulated Raman scattering. (b) A typical setup of pump-probe microscopy in which a modulation transfer scheme is utilized. (c) Resolution of the collinear pump-probe imaging as a function of NA, plotted according to Eq. (1). (d) Schematic diagram of the collinear and crossed-beam pump-probe excitation. (e) Different designs to implement crossed-beam pump-probe microscopy.

An alternative would be to use low-NA lenses. Low-NA lenses can achieve a large FOV and can be miniaturized, but offer relatively poor resolution, especially in the axial direction (Fig. 1(c)). For example, when the NA is 0.7, $r_{FWHM} = 0.43\ \mu m$ and $z_{FWHM} = 1.8\ \mu m$; but when the NA reduces to 0.1, $r_{FWHM} = 3.0\ \mu m$ and $z_{FWHM} = 100\ \mu m$. One does not always need intracellular resolving power; in these applications a $3\ \mu m$ lateral resolution, which is still smaller than many cell types (e.g. neuron, red blood cell), may be acceptable, but $100\ \mu m$ would clearly not resolve them in the axial direction. To enhance the axial resolution of low-NA lenses, we propose a pump-probe imaging method that utilizes crossed beams (Fig. 1(d)). We want to highlight that this method is different from the dual-axis confocal [13-15], dual-axis OCT [16], and dual-view PA imaging [17], where there is only one excitation beam and the detection is performed from a different angle. It is also different from vTwINS [18], where a V-shaped PSF is created to perform volumetric two-photon fluorescence imaging.

## 2. Use crossed beams to improve axial resolution

In pump-probe imaging, the acquired signal scales with the product of the pump and probe intensities [2]. If one uses a low-NA objective lens in the standard collinear setup (Fig. 1(d), left), the focused beam will have an elongated profile and thus results in a low axial resolution. In the crossed-beam configuration (Fig. 1(d), right), the pump and the probe beam propagate along different directions. Although the individual beams still have elongated profiles, the overlapping region now has a much smaller size along the $z$ axis. To estimate the axial resolution in this dual-axes configuration, we can approximate the beams as a uniform cylinder of diameter $2\omega_0$, where $\omega_0$ is the waist radius. The intersection of the overlap region in the plane spanning the illumination beam is a rhombus with diagonals of length $\Delta r$ and $\Delta z$:

$$r_{FWHM} \propto \Delta r = \frac{2\omega_0}{\cos(\theta/2)}, \qquad z_{FWHM} \propto \Delta z = \frac{2\omega_0}{\sin(\theta/2)} \qquad (2)$$

Eq. (2) shows that the axial resolution is proportional to the waist of the focused Gaussian beam and improves with increasing crossing angle ($0 < \theta < 90°$).

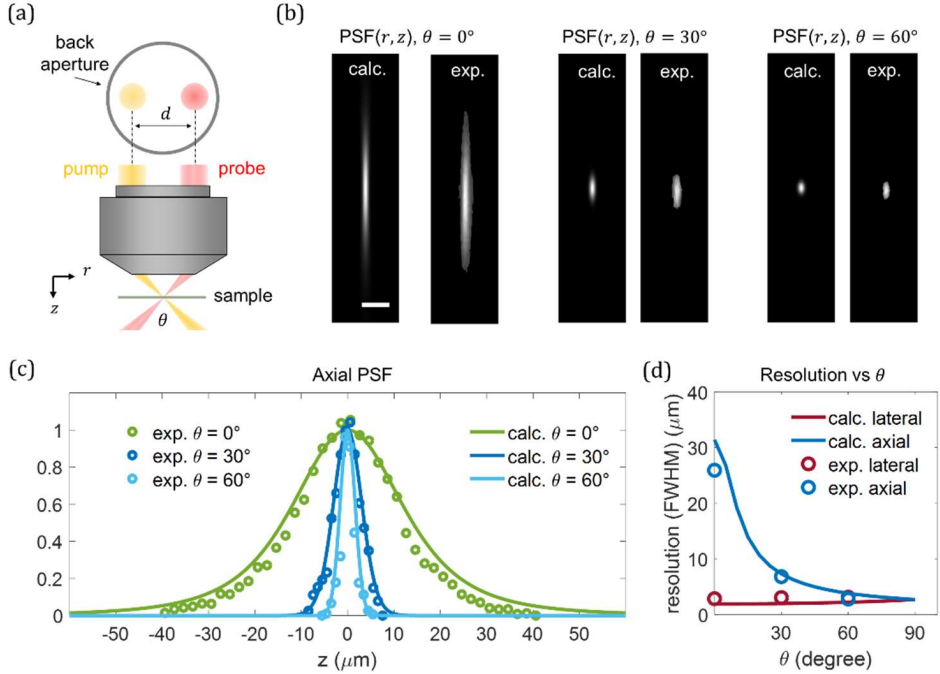

Fig. 2. (a) Configuration of the crossed-beam excitation in this experiment. The displacement between pump and probe is denoted by $d$. (b) Calculated (calc.) and measured (exp.) 2D PSF in the $rz$ plane. The scale bar is 15 μm. (c) Axial point-spread-function (PSF) of collinear ($\theta = 0°$) and crossed-beam ($\theta = 30°, 60°$) pump-probe microscopy. The solid lines and the circles are calculated and experimental results, respectively. (d) Calculated (solid lines) and measured (circles) resolution as a function of $\theta$, where $\theta = 0°$ corresponds to the collinear case.

We can also calculate the crossed-beam overlap region numerically assuming Gaussian beams with peak intensity $I_0$ and waist radius $\omega_0$ for pump and probe:

$$I(r, z; \lambda) = I_0 \left[ \frac{1}{1 + \left(\frac{z}{z_R}\right)^2} \right] exp \left[ \frac{-2r^2}{w_0^2 \left(1 + \left(\frac{z}{z_R}\right)^2\right)} \right] \qquad (3)$$

Here, $\lambda$ denotes the wavelength, and $r$, $z$ denote the lateral and axial direction, and $z_R = \pi \omega_0^2 / \lambda$ is the Rayleigh range. We calculate the point-spread-function (PSF) as the product of pump and probe beam intensity:

$$PSF = I_{pump} I_{probe} \qquad (4)$$

Figure 2 shows results from these calculations for $\lambda_1 = 0.71\,\mu m$, $\lambda_2 = 0.81\,\mu m$ (typical parameters for the melanin imaging experiments described below). For the waist radius, we used $\omega_0 = 2.3\,\mu m$, the value obtained in the experiment section *4.1*. We can see that in the collinear excitation (Fig. 2(b), left), the PSF of the low-NA lens is elongated in the $z$ direction, but in the cross-beam excitation (Fig. 2(b), middle and right), the PSF is more isotropic and the axial resolution is improved. Furthermore, we simulated how $\theta$ affects the imaging resolution (Figs. 2(c) and (d)), which shows that the axial resolution improves as $\theta$ approaches $90°$, whereas the lateral resolution drops by only a small amount. This feature can also be predicted by Eq. (2).

## 3. Experimental setup

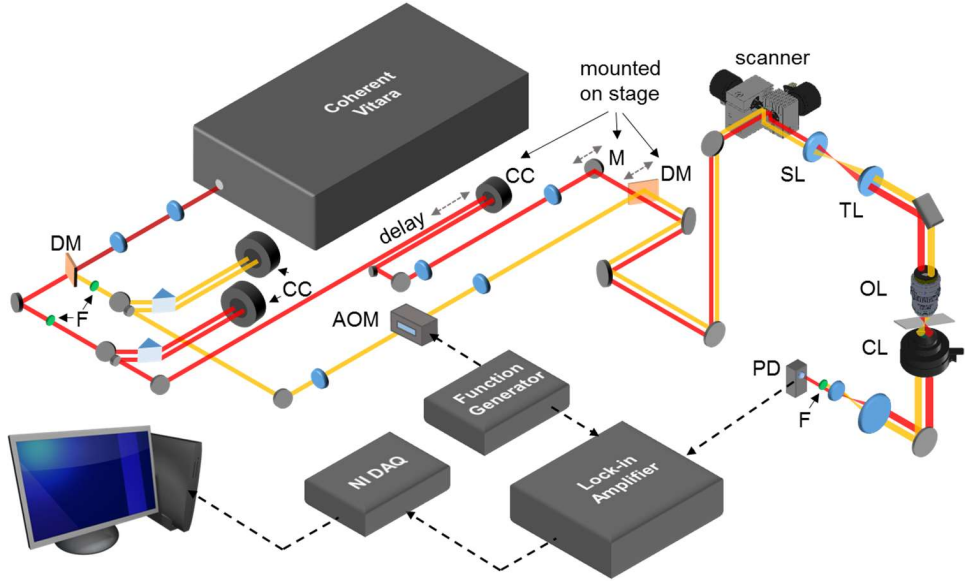

Fig. 3. Schematic diagram of the crossed-beam pump-probe microscope (CBPM). The pump beam (in orange) is modulated by an AOM at 1 MHz and the unmodulated probe beam (in red) passes through a delay line consisting of a corner cube mounted on a motorized stage. AOM: acousto-optic modulator, CC: corner cube, CL: condenser lens, DM: dichroic mirror, F: filter, OL: objective lens, PD: photodetector, SL: scanning lens, TL: tube lens.

There are three arrangements for crossed-beam excitation pump-probe imaging (see Fig. 1(e)): 1) using two lenses with a scanning sample, 2) using two lenses with beam scanning before illuminating the sample, or 3) using only one lens (either scanning the beam or scanning the sample). The first places no stringent requirements on the lenses, but requires sample scanning, which is generally slow. The second uses the same lenses, but requires post-objective beam scanning [9, 15], which is technically challenging because of space constraints. The third arrangement, using a single lens, is easier to implement, but the cross-angle is restricted by the NA of the lens. For this arrangement and using a high quality, high-NA objective lens, we do not expect crossed beams to offer an improvement over the full NA collinear case. However, as we demonstrate below, we observe an enhancement of the axial resolution of certain types of low-NA and low-quality lens. For this proof of principle experiment, we adopted the single-lens scheme as it allows us to use a standard laser-scanning microscope.

A schematic diagram of our crossed-beam pump-probe microscope (CBPM) is shown in Fig. 3. In this setup, the pump and probe beams are generated by splitting and filtering the output from a broadband ultrafast oscillator (Vitara UBB, Coherent Inc.). Both beams are dispersion-compensated in prism pulse compressors. The pump is modulated at 1 MHz by an acousto-optic modulator (AOM) (Model# TEF-110-50, Brimrose Corp.), and the probe is time delayed with respect to the pump in a delay line. The two beams are combined (parallel but offset from each other) by a dichroic mirror, and scanned by a XY galvo scanner (Cambridge Technology Inc.). By translating the dichroic mirror (DM) and the mirror (M) before it, we can adjust the beam separation of the pump and probe beams and therefore the intersection angle θ after the objective lens. For detection, the transmitted light is collected by a condenser lens (CL), the pump light is blocked by a bandpass filter (FB800-40, Thorlabs Inc.), and the probe light is directed onto a photodetector (PDA36A, Thorlabs Inc.). The PD signal is then analyzed in a lock-in amplifier (Model SR844, Stanford Research).

## 4. Experimental result

### 4.1 Characterization of the angle-dependent resolution of CBPM

In this experiment, we used a commercial objective lens (Olympus 20×/0.7 UPlanApo) and underfilled it with a narrow Gaussian beam (Fig. 2(a)). The effective NA is about 0.1, which gives a focused Gaussian beam with a waist radius of $\omega_0 = 2.3\ \mu m$, the value used in the simulation in section 2. Three angles ($\theta = 0°, 30°, 60°$) were tested by separating the pump and probe beams at different distances ($d = 0, 4.5mm, 8.6mm$), and the PSF was characterized by imaging gold nanostars [19]. Figure 2 (a) shows a schematic of the beams, panel (b) are 2D sections of the PSF in the in the $rz$ plane, acquired using the gold nanostars. Panel (c) shows the axial section of the PSF, and in (d) its axial and lateral width. Note that while we see a marked improvement in axial resolution using the crossed-beam arrangement, the lateral resolution expectedly drops from $0.9\mu m$ for the collinear, full-NA case to $1.3\mu m$ for the highest angle crossed-beam setup.

### 4.2 Compare the axial resolution enhancement of different lenses

Above we experimentally demonstrated the principles of CBPM using a high-quality, high-NA microscope objective. Here, we show that CBPM can improve the axial resolution of low-NA or non-imaging lenses compared to their full-NA resolving power.

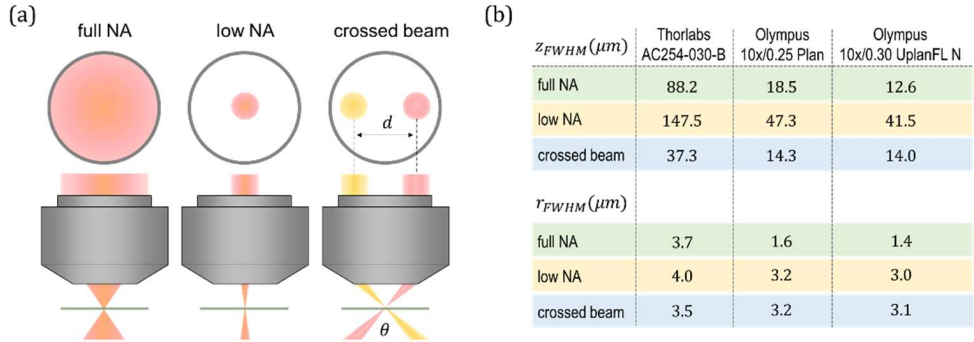

Fig. 4. (a) Schematic diagram of the collinear full NA, collinear low NA and crossed beam configuration used in this experiment. (b) The axial ($z_{FWHM}$) and lateral ($r_{FWHM}$) resolution of different lenses in all three configurations.

We characterized the resolution of three different lenses (Thorlabs AC254-030-B, Olympus 10×/0.25 Plan, and Olympus 10×/0.30 UPlanFL N) for three different configurations (Fig. 4(a)), again by imaging the gold nanostars. In the 'full NA' configuration, the beam size is expanded to overfill the back-aperture of the objectives, but not the 30mm achromatic lens. We did not overfill the achromat because it is not designed for high-quality imaging – after the beam size exceeds a certain value, aberrations will deteriorate the resolution. To determine this critical size we measured the resolution with a series of beam sizes and found it to be around 16 mm ($e^{-2}$ diameter). Therefore, for the 30 mm achromatic lens we expanded the beam size to this value ('full NA'). For the other two configurations ('low NA' and 'crossed beam'), the beam size was set to 4 mm for all lenses. Finally, in the 'crossed-beam' configuration the displacement between pump and probe was set to $d = 4.5\ mm$, which corresponds to different cross-angles in different lenses ($\theta \approx 17°, 13°$ and $16°$).

The measured axial and lateral resolution are shown in Fig. 4 (b). As expected, CBPM improves the axial resolution over the low-NA case for all three lenses. More importantly, crossed-beam matches (for the high-quality lens) and vastly improves (for the other lenses) the full-NA axial resolution even with small cross-angles. The latter case is most exciting, as it indicates that for low-NA and low-quality lenses the axial resolution in pump-probe imaging can be improved by simply displacing the two beams without modifying the microscope. Another interesting observation is that while CBPM generally reduces the lateral resolution due to the lower effective NA, in the case of the 30 mm achromatic lens CBPM actually improves this resolution. This is probably because of the large aberrations for full-NA operation of this lens (these lenses are not designed for large NA diffraction-limited imaging).

### 4.3 Application of CBPM

Above, we have characterized CBPM and demonstrated that crossed-beam pump-probe imaging can improve the axial resolution of low-NA and low-quality lenses. Here we demonstrate the imaging capability of CBPM in a biological sample and highlight the case where a long working distance is required.

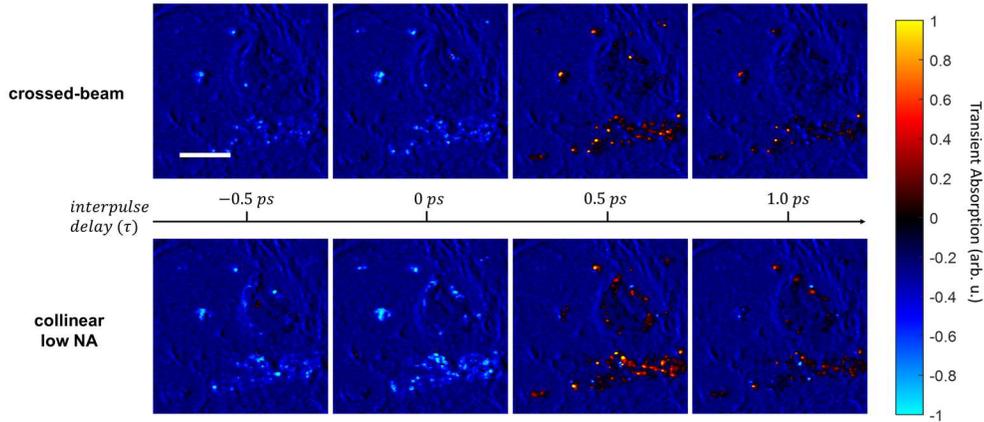

Fig. 5. Pump-probe imaging of a skin slice from a mouse. For the inter-pulse delay τ, positive values indicate the probe arrives after the pump pulse. Positive transient absorption values indicate nonlinear processes that lead to an increased attenuation of the probe beam (e.g. TPA, ESA), and negative values to a decreased attenuation (e.g. SRS, GSD). The scale bar is 60 μm.

First, we imaged a mounted skin biopsy slide (thickness about 10 μm) with the Olympus 10×/0.25 Plan objective that was underfilled with a beam size of 3 mm. The samples originated from a previous study on a malignant melanoma mouse model (the study had been conducted in accordance with an animal protocol approved by the Duke University Animal Care and Use Committee, see [20] for details). We imaged the sample using the crossed-beam and the low-NA collinear excitation configuration in the same region. The lateral and axial resolution of both configurations has been characterized in Fig. 4(e). The laser power used in this experiment is 1 mW for the pump (710 nm) and 0.3 mW for the probe (810 nm). Figure 5 shows the pump-probe images at different inter-pulse delays. In the color map, positive values (red, yellow) indicates nonlinear processes that increase probe absorption in the presence of the pump, such as two-photon absorption (TPA) and excited state absorption (ESA); negative values (blue) indicates processes that decrease probe absorption in the presence of the pump, such as ground state depletion (GSD). In Fig. 5, we can see that the crossed-beam configuration provides similar imaging contrast to the collinear configuration, with two minor differences. 1) The collinear pump-probe case shows slightly larger pixel intensities than the crossed-beam; the sample thickness is about the axial resolution of the crossed-beam configuration and hence, the overlap volume is slightly smaller. 2) Some pixels show positive values in the crossed-beam but not in the collinear configuration, which could be a sign of photodamage or thermal damage after repeated scanning.

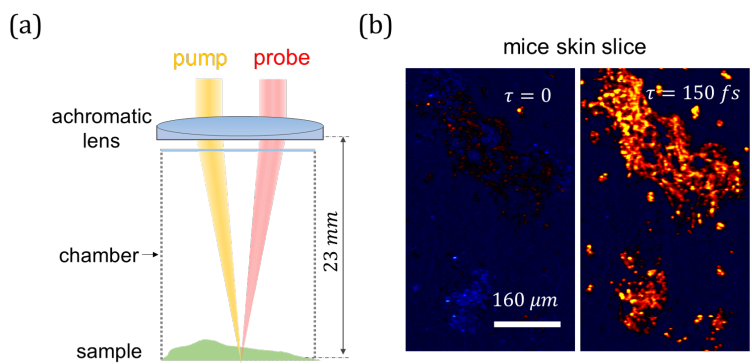

Fig. 6. (a) Schematic diagram of the long-working-distance CBPM using an achromatic doublet. The chamber serves to restrict the working distance. (d) Example image of fixed biological samples at two different time delays. The color map is the same as in Fig. 5.

Figure 6 illustrates an application where an extra-long working distance is required, such as imaging precious artwork, imaging vaporized materials, retina imaging through the eyeball, or imaging of samples that must be kept in a sealed chamber. In these cases, off-the-shelf achromatic lenses are preferred because of the cost, the size and the flexibility to provide a very large WD. Figure 6(a) shows the schematic diagram of our experiment: to mimic conditions requiring long WD, the sample is put in a chamber with a cover glass restricting access. Here, the objective lens is an achromatic doublet (Thorlabs AC254-030-B) – it has been characterized in the last section and provides a WD of 23 $mm$. Figure 6(b) shows the results when imaging the mouse skin sections at two different time delays, demonstrating a high-quality image at a large field-of-view.

### 5. Summary

We demonstrate that the crossed-beam excitation method in pump-probe microscopy can improve the axial resolution of low-NA and non-imaging, low-quality lenses. Such enhancement of axial resolution applies to most pump-probe imaging setups, including stimulated Raman scattering microscopy. CBMP can be implemented in a dual- or single-lens configuration; here, we use the second configuration for convenience. However, the dual-lens configuration is the better choice to achieve larger intersection angles and therefore better resolution enhancement. The significance of crossed-beam pump-probe microscopy (CBPM) is that it enables large field-of-view (FOV), long working-distances, and depth-resolved imaging by using achromatic lenses or parabolic mirrors, instead of specially designed objective lenses, and is scalable to a wide range of applications with different FOV and resolution requirements. Furthermore, CBPM minimizes dispersion and system complexity, which is beneficial for designing portable imaging devices.

## 6. Funding, acknowledgments, and disclosures

### 6.1 Funding

This project has been made possible in part by grant number 2019-198099 from the Chan Zuckerberg Initiative DAF, an advised fund of Silicon Valley Community Foundation (M.C.F.). This material is also based upon work supported by the National Science Foundation Division of Chemistry under Award No. CHE-1610975 (M.C.F.).

### 6.2 Acknowledgments

We thank Jin Yu, David Grass and Kevin Zhou for suggestions, Simone Degan, Pietro Strobbia and Shangguo Hou for providing samples, and Xuesong Li for providing drawing templates.

### 6.3 Disclosures

The authors declare no conflicts of interest.